\documentclass[aps,epsfig,graphics,twocolumn,floatfix,mathbbm,prl,superscriptaddress]{revtex4}
\usepackage{graphicx}

\begin{document}

\title[a]{Controlled Stark shifts in Er$^{3+}$-doped crystalline and
amorphous waveguides for quantum state storage}%

%

\author{Sara R. Hastings-Simon, Matthias U. Staudt, Mikael Afzelius}%
\address{Group of Applied Physics, University of Geneva, CH-1211 Geneva 4, Switzerland}%
\author{Pascal Baldi}%
\address{Laboratoire de Physique de la Mati\`{e}re Condens\'{e}e, Universit\'{e} de Nice-Sophia Antipolis,
Parc Valrose, 06108 Nice Cedex 2,  France}%
\author{Didier Jaccard}%
\address{D\'{e}partment de Physique de la Mati\`{e}re Condens\'{e}e, University of Geneva,
24 Quai Ernest Anserment, 1211 Geneva 4, Switzerland}%
\author{Wolfgang Tittel}%
\address{Institute for Quantum Information Science, University of Calgary,
2500 University Drive, NW Calgary, Alberta, Canada, T2N 1N4}%
\address{Group of Applied Physics, University of Geneva, CH-1211 Geneva 4, Switzerland}%
\author{Nicolas Gisin}%
\address{Group of Applied Physics, University of Geneva, CH-1211 Geneva 4, Switzerland}%

\begin{abstract}
We present measurements of the linear Stark effect on the
$^{4}$I$_{15/2} \rightarrow$ $^{4}$I$_{13/2}$ transition
 in an Er$^{3+}$-doped proton-exchanged LiNbO$_{3}$ crystalline
 waveguide and an Er$^{3+}$-doped silicate fiber.  The measurements were made
 using spectral hole burning techniques at temperatures below 4 K. We measured an effective Stark
 coefficient $(\Delta\mu_{e}\chi)/(h)=25\pm1$kHz/Vcm$^{-1}$ in the crystalline waveguide
 and $(\overline{\Delta\mu_{e}}\chi)/(h)=15\pm1$kHz/Vcm$^{-1}$ in the
 silicate fiber. These results confirm the potential of Erbium doped waveguides for quantum state storage based on
controlled reversible inhomogeneous broadening.
\end{abstract}
\maketitle

Because of their narrow homogenous line widths and corresponding
long optical coherence times at low temperatures, rare-earth doped
crystals (RE doped crystals) have been the subject of much
investigation for photon-echo based optical data storage and data
processing \cite{photonecho}. Recently, there has been additional
interest in RE doped crystals, spurred by proposals to extend
their use to applications in the fast developing field of quantum
communication and computation \cite{chuang00}.  In particular, RE
doped crystals are a promising material for the realization of a
reversible transfer of quantum states between photons and atoms.
Such a quantum memory represents a basic building block for the
so-called quantum repeater \cite{Briegel98}, which allows the
extension of quantum communications schemes such as quantum
cryptography \cite{gisin02} to unlimited distances.

A original proposal for a quantum, as well as classical, memory in
solid state material is based on controlled reversible
inhomogeneous broadening (CRIB) of a narrow, spectrally isolated
absorption line \cite{memory}.  This protocol requires an atomic
ensemble with a large optical depth, a long optical coherence
time, and the ability to inhomogeneously broaden a single
absorption line in the ensemble in a controlled and reversible
way.

Erbium doped waveguides are promising materials for the
realization of CRIB, as interaction lengths ranging from many cm
(in the case of crystalline waveguides), to hundreds of meters or
more (in the case of amorphous optical fibers) can easily be
achieved, allowing for large absorption even at low doping
concentrations. The 1.5 $\mu$m, $^{4}$I$_{15/2} \rightarrow$
$^{4}$I$_{13/2}$, transition in Er$^{3+}$ is well matched to
standard telecommunication fiber, which would allow simple
interfacing of such a memory in a fiber communication network.  In
addition, recent investigations show that coherence times on the
scale of $\mu$s can be obtained in Er$^{3+}$ doped silicate fibers
\cite{macfarlane,staudt} , and on the order of ms in Er$^{3+}$
doped crystals \cite{Bottger}. In this article we investigate the
linear Stark effect for the realization of CRIB in two different
Erbium doped waveguides, a crystalline LiNbO$_{3}$ waveguide and a
silicate optical fiber. Our results confirm the potential of
Erbium doped waveguides for CRIB-based data storage.


In the presence of an external DC electric field, the energy
levels of an atom with a permanent dipole moment are shifted; this
phenomenon is known as the linear DC Stark effect. If the dipole
moments are different for different electronic levels this shift
leads to a shift in the associated optical transition frequency.
The linear DC Stark effect can be observed in RE doped solids,
where the dipole moment in the ion is induced by local electric
fields. It has been measured for the $^{4}$I$_{15/2} \rightarrow$
$^{2}$H$_{11/2}$ transition in Er$^{3+}$:LiNbO$_{3}$
\cite{ErLiNbO3} and Er$^{3+}$,Mg$^{2+}$:LiNbO$_{3}$
\cite{ErMgLiNbO3} crystals, using a differential technique
\cite{differential}. In addition the effect has been measured in
RE doped bulk crystals such as Eu$^{3+}$:YAlO$_{3}$
\cite{EuYAl,Eu:YAl03_2}, Pr$^{3+}$:YSiO$_{5}$
\cite{Pr:YSi05andEUPr:YAl03}, Pr$^{3+}$:YAlO$_{3}$ \cite{Pr:YAl03}
and Pr$^{3+}$:LaF$_{3}$ \cite{PR:LAF} as well as in Sm$^{2+}$
doped borate bulk glass \cite{Sm}, an amorphous material. In order
to achieve high sensitivity in the latter measurements, spectral
hole burning techniques \cite{holeburning} are used to isolate the
effect of an electric field on a single homogenous line within an
inhomogeneously broadened ensemble of atoms. To the best of our
knowledge no measurements have been reported on RE doped
waveguides, neither crystalline nor amorphous.

We consider the linear Stark effect where the transition
frequency, $\omega$, of an ion is shifted by the electric field
according to the formula,

\begin{equation}
\Delta\omega = \frac{\Delta\mu_{e}\chi E}{\hbar}\cos\theta.
 \label{stark}
\end{equation}

Here $\Delta\mu_{e}$ is the difference between the permanent
electric dipole moments of the two states connected by the
transition, $E$ is the applied DC electric field,
$\chi=(\epsilon+2)/3$ is the Lorentz correction factor, $\epsilon$
is the dielectric constant of the sample, and $\theta$ is the
angle between the vectors $\vec{\Delta\mu_{e}}$ and $\vec{E}$.

In a crystal, because of the ordered structure of the crystallin
lattice, the dipole moments are aligned along a set of one or more
well defined directions.  The symmetry of the crystal along with
the site symmetry at the rare-earth-ion position determines the
number of directions. All ions with $\vec{\Delta\mu_{e}}$ aligned
along the same direction experience the same shift in their
resonance frequency with an applied electric field and thus the
application of a DC electric field leads to a shift or splitting
of the spectral hole, depending on the number of possible
directions. The projection of $\Delta\mu_{e}$ between the two
states, along the direction of the applied electric field, can be
determined by measuring the shift of the hole as a function of the
electric field, as given by equation \ref{stark}.

In an amorphous material, such as an optical fiber, the dipole
moments are randomly orientated, such that the projection of
$\vec{\Delta\mu_{e}}$ along the direction of the applied electric
field varies continuously for different ions in the solid.
Therefore, rather than a shift or splitting of the spectral hole,
one observes a broadening as the absorption frequencies shift by
different amounts, in different directions. Assuming a Lorentzian
spectral hole, one can derive the shape of the hole for a given
applied electric field. Here we make the following simplifying
assumptions as proposed by Bogner et al. \cite{Bogner}. First, the
dipole moments are randomly orientated in the solid. Second, there
is a distribution of magnitudes of $\Delta\mu_{e}$ which can be
described by a Maxwell Distribution:

\begin{equation}
  g(\Delta\mu_{e})=\frac{4}{\pi^{1/2}{(\overline{\Delta\mu_{e}})}^{3}}(\Delta\mu_{e})^{2}
  \exp\left[-\left(\frac{\Delta\mu_{e}}{\overline{\Delta\mu_{e}}}\right)^{2}\right].
 \label{dipoledistribution}
 \end{equation}

Here $\Delta\mu_{e}$ is the difference in dipole moment between
the two states, and $\overline{\Delta\mu_{e}}$ is the most likely
value of this difference.  Note that the width of the distribution
and the value of $\overline{\Delta\mu_{e}}$ are not independent.

With these assumptions the shape of the spectral hole,
\textit{h(x)}, under application of an electric field, is given by
\cite{Kador}

\begin{eqnarray}
h(x)=\frac{2}{\pi^{3/2}\overline{f}^{3}}
\bigg\{\int_{0}^{\sqrt{1+x^{2}}}
 f\exp\left[-\left(\frac{f}{\overline{f}}\right)^{2}\right]*
 \nonumber
\\
 \arctan\left(\frac{2f}{1-f^{2}+x^{2}}\right)
 \,df+
\\
\int_{\sqrt{1+x^{2}}}^{\infty}
f\exp\left[-\left(\frac{f}{\overline{f}}\right)^{2}\right]*
\nonumber
\\
 \left[\pi+\arctan\left(\frac{2f}{1-f^{2}+x^{2}}\right)\right]
\,df \bigg\}
 \nonumber
 \label{hole}
 \end{eqnarray}

with

 \begin{equation}
  \overline{f}=\frac{\overline{\Delta\mu_{e}}\chi E}{\hbar\gamma} \ \
  \ \ \ x=\frac{(\omega-\omega_{C})}{\gamma}.
 \label{fvariable}
 \end{equation}

Here $\omega$ is the laser frequency, $\omega_{C}$ is the central
transition frequency, $\gamma$ is the full width at half maximum
(FWHM) of the spectral hole without applied electric field, and
the other variables are defined as in equation \ref{stark}.


In our investigation of the Stark effect in Er$^{3+}$ doped
waveguides we used an monochromatic laser to excite Er$^{3+}$ ions
from the $^4I_{15/2}$ ground state to the $^4I_{13/2}$ excited
state, thereby creating a spectral hole at the frequency of the
laser. We then decreased the laser intensity and scanned the
frequency around the initial burning frequency three times while
measuring the transmitted light as a function of time. Together
with an independent time-to-frequency calibration of the laser
scan, this yields three spectral hole profiles. During the first
scan there was no applied electric field, for the second scan we
applied an electric field across the sample, as described below.
Finally, for the third scan, we switched off the electric field to
demonstrate the reversibility of the Stark induced line shift.

The first sample, a single domain, z-cut, proton-exchanged
LiNbO$_{3}$ waveguide (dimensions 1 mm thick, 6 mm long, 6 mm
wide) is doped with Er$^{3+}$ (0.5 at \%), and MgO (5 mol \%). We
mounted the waveguide between two metal electrodes spaced by 1 mm
and cooled it to 3.8 K using a pulse tube cooler (VeriCold
Technologies). Light, linearly polarized along the crystal C$_{3}$
symmetry axis, was injected and collected with standard optical
fibers mounted on two xyz positioners (Attocube Systems) and
aligned with the input and output of the waveguide. The C$_{3}$
symmetry axis was oriented parallel to the applied electric field
and perpendicular to the light propagation direction. In addition
we applied a magnetic field of approximately 0.7 T parallel to the
C$_{3}$ symmetry axis to reduce the width of the spectral hole
\cite{magneticfield}. We performed hole burning experiments as
described above, with applied voltages ranging from -200 to 200 V,
using a scannable cw external cavity diode laser (Nettest, Tunics
Plus) at a wavelength of 1531.00 nm. To measure the spectral hole
we scanned the laser over 1.2 GHz in 500 $\mu$s and measured the
transmitted light with a photodiode (NewFocus, mod. 2011).

\begin{figure}
\centering
 \includegraphics[width=.4\textwidth]{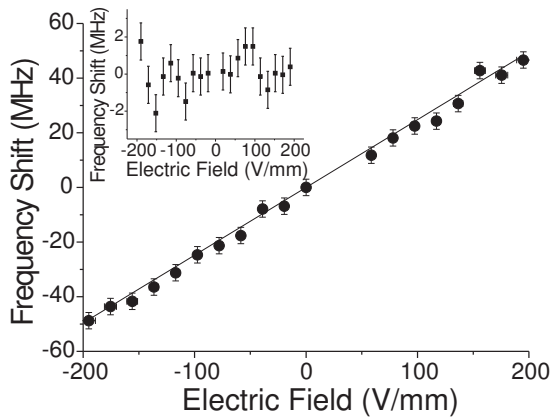}
  \caption{The shift of the spectral hole central frequency
as a function of electric field in the Erbium doped LiNbO$_{3}$
waveguide.  The line is a linear fit to the data, which passes
through the point (0,0), and gives an effective dipole moment
difference of $(\Delta\mu_{e}\chi)/(h)=25\pm1$kHz/Vcm$^{-1}$. The
inset graph shows the frequency shift of the hole after the
applied field is turned off.}

  \label{crystal}
  \end{figure}

The frequency shift of the spectral hole, relative to the zero
field position, as a function of electric field is plotted in
figure \ref{crystal}. The line is a linear fit to the data which
passes through the point (0,0). The dipole moment difference can
be calculated from the slope of the line according to equation
\ref{stark}.  We find an effective dipole moment difference
$(\Delta\mu_{e}\chi)/(h)=25\pm1$kHz/Vcm$^{-1}$. The inset graph in
figure \ref{crystal} shows the position of the spectral hole
measured after switching off the electric field for each applied
field; the spectral hole returns to the initial zero field
position. This, plus the opposite directional shifts that were
observed with positive and negative voltages, confirms the
reversibility of the Stark effect.

The observed single-frequency shift of the absorption line, under
application of an electric field in the direction of the C$_{3}$
symmetry axis, reflects the lack of inversion symmetry in the
LiNbO$_{3}$ crystal.  This property allows for a single dipole
moment direction, in contrast to observations of a "pseudo Stark
splitting" \cite{pss} in centrosymmetric crystals, where the
inversion symmetry results in dipole moments pointing in opposite
directions.

The second sample, a SiO$_2$ fiber (Ino, ER 407 Fiber) is doped as
follows: Er 0.07 (at \%), Al 2.65 (at \%), Ge (3.62 at \%). The
fiber (length 65 cm) was coiled with a diameter of 4 cm and
sandwiched between two round metal electrodes spaced by 0.3 mm. We
placed the sample in a $^{3}$He/$^{4}$He dilution refrigerator and
cooled it to 50 mK. As the spectral hole width is sufficiently
small at this temperature, application of a magnetic field was not
necessary. We performed spectral hole burning experiments as
described above, with applied voltages ranging from 0 to 120 V,
using a cw external cavity diode laser (Toptica, DL 100) at a
wavelength of 1531 nm. To measure the spectral hole we scanned the
laser over 400 MHz in 800 $\mu$s. The transmission was measured
with a photodiode (NewFocus, mod. 2011).

\begin{figure}
\centering
 \includegraphics[width=.4\textwidth]{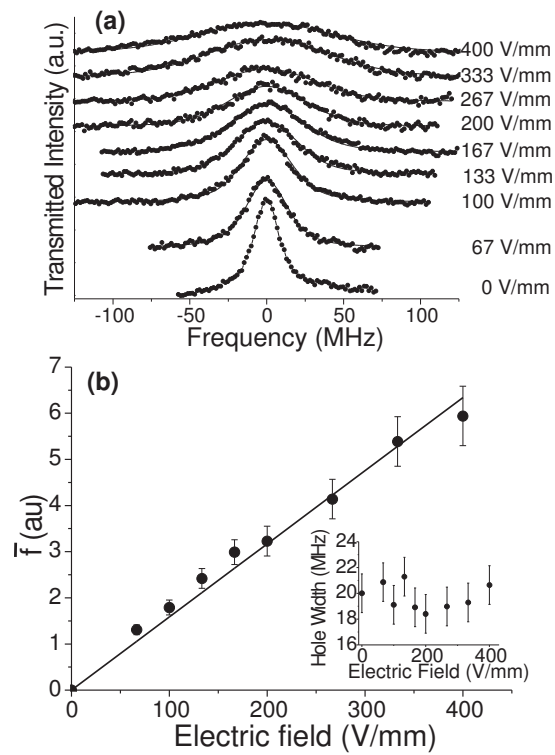}
  \caption{Spectral holes in Er$^{3+}$ doped silicate telecom fiber, as a function of applied electric field. Figure (a) shows the
spectral holes (transmitted intensity as a function of frequency)
with different applied electric fields, the holes are shifted
vertically for clarity. The points are measured data and the lines
are a fit to equation 3. Figure (b) is a plot of the fitted
$\overline{f}$ values from figure (a).  The line is a linear fit
to the data, which passes through the point (0,0), and gives  an
effective dipole moment difference
$(\overline{\Delta\mu_{e}}\chi)/(h)=15\pm1$kHz/Vcm$^{-1}$. The
inset graph in figure (b) shows the width of the spectral hole
(FWHM) measured after the applied field is turned off.}
  \label{fiber}
  \end{figure}

The spectral hole for each applied electric field is plotted in
figure \ref{fiber}a. The points are the measured data and the
lines are fits using equation \ref{hole}. We find a close
agreement between the predicted and observed hole shapes. The
values of $f$ obtained from the fits for each applied electric
field are plotted in figure \ref{fiber}b. The line is a linear fit
to the data which passes through (0,0). The dipole moment
difference can be calculated from the slope of the line according
to equation \ref{fvariable}. We find an effective dipole moment
difference
$(\overline{\Delta\mu_{e}}\chi)/(h)=15\pm1$kHz/Vcm$^{-1}$. The
inset graph in figure \ref{fiber}b shows the width of the spectral
hole (FWHM) measured after switching off the electric field for
each applied field. The spectral hole width returns to the initial
zero field value, which confirms that the broadening of the
spectral hole can be removed.

Let us now discuss our results in view of the controlled
reversible inhomogeneous broadening required for the quantum
memory proposal \cite{memory}. The broadening must be controllable
in such a way that, starting with an ensemble of ions that absorb
at $\omega_{0}$, the transition frequency of the ion can be
shifted to $\omega_{0}+\omega_{b}$ and then, at a later time,
shifted again to $\omega_{0}-\omega_{b}$, where $\omega_{b}$
varies continuously over the necessary bandwidth. For the storage
of light pulses with a duration of 10ns, for example, the required
frequency bandwidth is on the order of 100 MHz. In the crystal
this broadening could be achieved by applying a position dependent
electric field along the crystal, ranging from +2 kV cm$^{-1}$ to
--2 kV cm$^{-1}$. In the fiber, the application of a spatially
homogeneous electric field of approximately 3.5 kV cm$^{-1}$ is
sufficient as it directly leads to a broadening of the spectral
hole. In both the fiber and the crystal, switching the polarity of
the applied field should reverse the broadening in the required
way.


In conclusion, we have presented measurements of the linear DC
Stark effect in an Er$^{3+}$-doped proton-exchanged LiNbO$_{3}$
waveguide, and an Er$^{3+}$ doped silicate telecom fiber. We found
effective dipole moment differences of
$(\Delta\mu_{e}\chi)/(h)=25\pm1$kHz/Vcm$^{-1}$ and
$(\overline{\Delta\mu_{e}}\chi)/(h)=15\pm1$kHz/Vcm$^{-1}$,
respectively. These measurements demonstrate the suitability of
Er$^{3+}$ doped waveguides for solid state quantum memory
protocols based on controlled reversible inhomogeneous broadening.

\section{Acknowledgments}

We would like to thank M. Nilsson for useful discussions.
Technical Support by C. Barreiro and J.-D. Gauthier is
acknowledged. This work was supported by the Swiss NCCR Quantum
Photonics and the European Commission under the Integrated Project
Qubit Applications (QAP) funded by the IST directorate as Contract
Number 015848. Additionally, M.A. acknowledges financial support
from the Swedish Research Council.

\end{document}